\documentclass[twocolumn,amsmath,aps,nofootinbib,superscriptaddress]{revtex4}
\usepackage{graphicx}
\usepackage{subfigure}
\usepackage{bm}
\usepackage{amsmath}
\usepackage{xcolor}
\usepackage{amssymb}
\usepackage{mathrsfs}

\makeatletter

\newcommand{\Rmnum}[1]{\expandafter\@slowromancap\romannumeral #1@}
\makeatother

\DeclareMathOperator{\sech}{sech}

\begin{document}

\title{Cosmological Perturbations of Extreme Axion in the Radiation Era}

\author{Ui-Han Zhang}
\affiliation{Department of Physics, National Taiwan University, Taipei 10617, Taiwan}

\author{Tzihong Chiueh}
\email{chiuehth@phys.ntu.edu.tw}
\affiliation{Department of Physics, National Taiwan University, Taipei 10617, Taiwan}
\affiliation{National Center for Theoretical Sciences, National Taiwan University, Taipei 10617, Taiwan}

\begin{abstract}
\label{Abstract}
Sub-horizon perturbations under the extreme initial condition of the axion model are investigated, where initial axion angles start near the potential maximum. This work focuses on a few new features found in the extreme axion model but absent in the free-particle model.  A particularly novel new feature is the spectral excess relative to the CDM model in some wave number range, where the excess may be so large that landscapes of high-redshift universe beyond $z=10$ can be significantly altered.  For axions of particle mass $10^{-22}$ eV, this range of wave number corresponds to first galaxies of few times $10^9-10^{10} M_\odot$.  We demonstrate that sub-horizon perturbations are accurately described by Mathieu's equation and subject to parametric instability, which explains this novel feature.  Actually the axion model is not a special one; perturbations in a wide range of scalar field models can share the similar characteristic.
\end{abstract}

\maketitle
\section{Introduction}
\label{sec: Introduction}

Scalar fields as dark matter candidates have a long history of development \cite{DR1993, Sin1994, JS1994, DRA1995, LK1996, Hu2000, GM2000, MGN2000, MU2000, SW2000}, but most works were addressing the matter-dominated era where comparisons with observations can be made.   While these models have some degrees of freedom to accommodate a suite of observational oddities, it is inevitable that they must introduce one or more energy or mass scales, in sharp contrast to the CDM model, which is extremely insensitive to the particle mass.  For example, wave dark matter ($\psi$DM)\cite{Schive2014},  ultra-light bosonic dark matter \cite{WooChiueh2009} or fuzzy dark matter \cite{Hu2000} introduces one boson mass $m$, which is found to be around $10^{-22}$ eV to explain kilo-parsec-scale core structures in dwarf spheroidal galaxies \cite{Schive2014, CSC2017}. If $m \gg 10^{-22}$ eV, $\psi$ dark matter becomes indistinguishable from CDM observationally.  The axion model, being a nonlinear field model, introduces a second energy scale $f$, in addition to $m$, where $f$ is the axion decay constant that is above the GUT scale to explain the cosmic background dark matter mass density to be so close to the cosmic critical density in a non-QCD axion model involving the dark sector\cite{ChiueharXiv2014, DM2017}.  Indeed, recent developments of string theories also favor extremely light axions with a large axion decay constant $f$ much greater than the electroweak scale \cite{SW2006, ADDKM2010, HOTW2017, Diez-TejedorMarsharXiv2017}.

However, there is one more free degree of freedom, i.e., the initial field amplitude, which is a dimensionless parameter not present in the field Lagrangian but is able to control the solution.   Whether the initial field is located in a linear regime or in a nonlinear regime may make a difference in the solution space and affects the observable. In the context of cosmology, as the universe expands the field amplitude quickly decreases due to Hubble friction, and soon the field samples only the quadratic part of the potential to become free particle.   Hence the free-particle model ($\psi$DM) is the ultimate time asymptotic attractor for the axion model and for many other nonlinear scalar field models.  One therefore hopes that the Hubble friction may erase the memory of the initial condition, and the solution converges to the free-particle solution.

We therefore studied linear perturbations of the free-particle model in the radiation dominant era in the previous work \cite{ZhangChiueharXiv2017} (Paper (\Rmnum{1})).  Four phases of evolution are identified.  Central to the four phases is the critical wavenumber $k_c$, for which the mode enters horizon when the horizon size equals the Compton wavelength.  This critical wavenumber lies at the boundary of the four phases and gives rise to a sharp spectral transition.   We have also numerically investigated perturbations of the axion model to investigate the attractor aspect of the problem, and indeed found that the time-asymptotic solution depends very weakly on the initial angles, except  when the axion field starts from very close to the top of the field potential, a highly nonlinear initial field.   In such an exceptional case, the perturbation begins to behave quite unexpectedly from when the field starts elsewhere.  We call this singular case the extreme axion model.  This narrow window of new degree of freedom is interesting, and may allows for accommodating the tension concerning the particle mass of $\psi$DM determined by the high-redshift Lyman-$\alpha$ forests \cite{APYNB2017,IVHBB2017} and by the flat cores of nearby dwarf spheroid galaxies \cite{CSC2017, LM2014, CS2016}.  In this work, we follow up this finding of Paper (\Rmnum{1}) for the extreme axion model and analyze the perturbation evolution in details.  Particular emphasis is placed on sub-horizon modes after the onset of mass oscillation, as it holds the key to the unexpected.  The analysis developed in this work can be extended to other nonlinear scalar field models with a finite potential barrier.

For the fiducial boson mass $m$ as small as $10^{-22}$ eV, the particle number density is extremely high, yielding a critical temperature so high that any conceivable background temperature is way below the critical.  These bosons therefore form a Bose-Einstein condensate and many-particle wave functions collapse to a single wave function.  To acquire phase coherence for many-particle wave functions, nonlinearity is essential to couple these wave functions and locks their phases.   The nonlinearity of the scalar field for a Bose-Einstein condensate is just a manifestation of the microscopic two-body scattering that correlate wave functions.  To the leading order of interactions, the simplest S-wave scattering of a dilute boson gas results in the Gross-Pitaevskii equation as an effective macroscopic theory \cite{Gross1961, Pitaevskii1961} via the well-known Bogoliubov's reduction formalism in the non-relativistic limit \cite{Bogoliubov1947}.  We will show that a general class of relativistic scalar field models can be reduced to the Gross-Pitaevskii equation to the leading-order nonlinearity in the non-relativistic limit, and the axion model is no exception.   Therefore the axion model gains a microscopic support for being a Bose-Einstein condensate, and this finding will be elaborated later.

Recently there have been numerical works attempting to address axion perturbations \cite{LCarXiv2017}, partly motivated by string theories \cite{SW2006, ADDKM2010, Diez-TejedorMarsharXiv2017} and partly by the emerging interest in wave dark matter\cite{Schive2014, CSC2017, APYNB2017,IVHBB2017, HOTW2017, MS2014, SchivePRL2014, MP2015, Marsh2016}.  The difficulty of computing axion perturbations arises from that the equation demands high numerical accuracy to solve, as it must stably track two near-by-frequency oscillations for thousands of, or even much more, periods to determine the precise relative phase shift between the two oscillations.  It is therefore essential that numerical results have a support from detailed analyses of the solution.  One of the aims of this work is to serve for such a purpose.   We find for the free-particle case, as in Paper (\Rmnum{1}), all numerical works largely agree.  For the extreme axion case, we find the numerical results of other works \cite{Diez-TejedorMarsharXiv2017, LCarXiv2017} deviate from the result of this work.   It remains to be seen at what numerical bottleneck these differences arise.

This paper is structured as follows.  In Sec. (\ref{sec: Review on the Peculiar Features of the Extreme Axion}), we briefly review the previous work and pose problems pertinent to the three unexpected features to be understood.  Sec. (\ref{sec: Exponential Growth of delta_theta}) addresses the first feature.  The remaining two features require a new mechanism involving the parametric drive and amplification, and we elaborate this mechanism in Sec. (\ref{sec: Parametric Instability}).  The treatment can generalized to other scalar field models as shown in Sec. (\ref{sec: Numerical Solution and General Nonlinear Model}).  The matter power spectrum of the extreme axion model is also shown in this section.  In Sec. (\ref{sec: Axion connection to Gross-Pitaevskii Equation}), we make contact of the axion model to the Gross-Pitaevskii equation and also discuss the concern about quantum tunneling when the initial field assumes a classically unstable value.   We conclude this work in Sec. (\ref{sec: Conclusion}).  The particle mass dependence of our results is discussed in Appendix.  This work is confined to the radiation-dominant era unless otherwise mentioned, the fiducial boson mass is chosen $10^{-22}$ eV, and standard cosmological parameters of the concordance model are adopted, i.e., $H_0 = 70 km/sec/Mpc$, $\Omega_{DM}=0.24$, $\Omega_b=0.06$.  We also set the speed of light $c$ and the Planck constant $\hbar$ equal to $1$.  Throughout the analysis we adopt the Newtonian gauge for perturbations.

\section{Review on Unique Features of the Extreme Axion Model}
\label{sec: Review on the Peculiar Features of the Extreme Axion}

In the following, we shall first adopt the axion model with the $m^2a^2(1-\cos(\theta))$ field potential as a working template, and in a later section we will extend the analysis to more general nonlinear potentials.  The equations of motion for the axion background field $\theta$ and the perturbed field $\delta\theta$ are respectively
\begin{equation}
\label{equ: background field equation}
\theta^{''} + 2H\theta^{'} + m^2 a^2 \sin(\theta) = 0,
\end{equation}
and
\begin{equation}
\label{equ: perturbed field equation}
\delta\theta^{''} + 2H\delta\theta^{'} + [k^2 + m^2a^2\cos(\theta)]\delta\theta = 4\phi^{'}\theta^{'} - 2m^2a^2\sin(\theta)\phi,
\end{equation}
where the prime denotes the derivative with respect to the conformal time $\tau(\equiv\int_0 dt/a\propto a)$, $a$ is the scaling factor, $H(\equiv da/dt\propto a^{-1})$ the conformal Hubble parameter, and $k$ the wavenumber. The right-hand side of the perturbed field equation is the source $Sr$, which contains the metric perturbation $\phi$ and is contributed from perturbations of all species through the Poisson equation.  The metric perturbation is dominated by the photon perturbation in the radiation era, and to a good approximation we can regarded $\phi$ as independent of $\delta\theta$ till near radiation-matter equality.  This approximation is called the passive evolution in Paper (\Rmnum{1}).  The full treatment of $\phi$ must include additionally the dark matter, baryons and neutrinos, as elucidated in Paper (\Rmnum{1}), which shows that passive evolution provides a good approximation before the epoch of radiation-matter equality.

We need also define one more quantity, the dimensionless gauge-covariant energy density of the perturbed field
\begin{equation}
\label{equ: definition of Delta psi}
\Delta_\theta \equiv {{\theta^{'}\delta\theta^{'} + m^2a^2\sin(\theta)\delta\theta - (\theta^{'})^{2}\phi +3H\theta^{'}\delta\theta}\over{(1/2)(\theta^{'})^{2} + m^2a^2[1 - \cos(\theta)]}},
\end{equation}
which is the normalized physical energy density perturbation and the denominator is the background field denisty $\epsilon_\theta$.

A short summary of Paper (\Rmnum{1}) is in order.   In the small $\theta$ limit, which most parts of Paper (\Rmnum{1}) addresses, is the free-particle limit as the field nonlinearity vanishes.  In this limit, we have well-defined four phases, (a) before mass oscillation and superhorizon, (b) before mass oscillation and subhorizon, (c) after mass oscillation and super-horizon, and (d) after mass oscillation and subhorizon.  Long wave go through phases (a),(c) and (d), but short waves through (a),(b) and (d).  The division of long and short waves is the critical wavenumber $k=k_c$, where $k_c=ma=2H$, for which all terms on the left-hand side of Eq. (\ref{equ: perturbed field equation}) are equally important and mass oscillation just begins to set in.   In phase (a), $\Delta_\theta$ grows as $a^6$, phase (b) as $a^2\cos((ka/\sqrt{3})-\chi_1)$, phase (c) as $a^2$ and phase (d) as $\sin((k^2/2maH)\ln(a)-\chi_2)$, where $\chi_1$ and $\chi_2$ are oscillation phases associated with photon and matter wave, respectively.

\begin{figure}
\includegraphics[scale=0.35, angle=270]{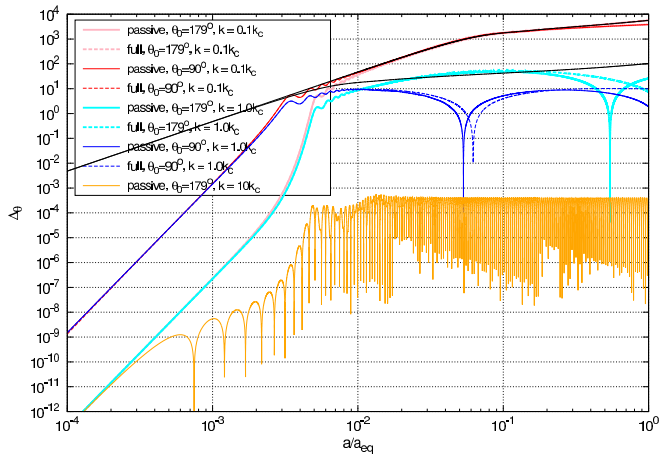}
\caption{Evolution for passive and full treatment of $\Delta_\theta$ with different inital angles ($\theta_0 = 90^{o}$ and $179^{o}$) for wavenumbers $k=0.1k_c$, $1.0k_c$ and $10k_c$. Particle mass is assumed to be $10^{-22}$ eV. Also plotted in black solid lines are the CDM model of $k=k_c$ and $0.1 k_c$ for references. This figure demonstrates the three features for the extreme axion model, particularly for the $k=k_c$ case discussed in Sec. (\ref{sec: Review on the Peculiar Features of the Extreme Axion}). }
\label{fig: compare_passive_full_axion_90_179}
\end{figure}

When the initial angle $\theta_0$ is not small, the $\theta_0=\pi/2$ case is of very little difference from the free-particle model found in Paper (\Rmnum{1}).   In fact, Fig. (\ref{fig: compare_passive_full_axion_90_179}) sums up nicely the features to be discussed, which illustrates the evolution of $\Delta_\theta(a)$ for long, medium and short waves with initial angles $\theta_0 = 179^{o}$ and $90^{o}$.  When comparing  $\theta_0 = 179^{o}$ and $90^{o}$ cases, the first feature common to all wave numbers is (1) a steep rise in amplitude at the onset of mass oscillation for $\theta_0 =179^{o}$ not present for $\theta_0=90^{o}$.  A second feature of the $\theta_0=179^{o}$ case is (2) a substantially longer duration of the first half cycle of matter-wave oscillation for some $k\sim k_c$ than that of the $\theta_0 = \pi/2$ case for the same $k$.  Associated with the second feature is a third feature that (3) the perturbation amplitude of the $\theta_0=179^{o}$ case is higher than that of the CDM model during a certain period also for some $k\sim k_c$, which has never been observed in the free-particle model.   Clearly, these three features are not caused by ordinary nonlinear mass oscillation of $\theta$, but associated with the extreme condition where $\theta_0 \to \pi$.  Note that the second and third features do not show up prominently for $k\ll k_c$ and $k\gg k_c$ in Fig. (\ref{fig: compare_passive_full_axion_90_179}).  This requires an explanation.

Finally, Fig. (\ref{fig: compare_passive_full_axion_90_179}) demonstrates that the passive evolution approximates the evolution of full treatment quite well till near the radiation-matter equality.  The focus of our analysis in this work is placed upon after the onset of mass oscillation but still far away from the epoch of radiation-matter equality.  Hence, passive evolution provides a fair simplification for understanding the above three features; however, our numerical solutions will include the full treatment.

\section{Abrupt Growth of $\delta\theta$}
\label{sec: Exponential Growth of delta_theta}

Prior to the onset of mass oscillation, the perturbation grows as $a^6$, corresponding to the earliest phase (a) in the evolution.  When $\theta_0$ is near the top of the field potential, it delays the mass oscillation, and substantial delay makes the friction $2H\delta\theta^{'}$ negligible in Eq. (\ref{equ: perturbed field equation}) at the onset of mass oscillation.  This creates an almost frictionless background for perturbed field dynamics. This rapid growth occurs only in a short time when $\theta$ first rolls down from the potential top.  The duration of exponential growth is independent of the exact location of the initial angle $\theta_0$ from the potential top as long as $\theta_0$ is close to the top.  More importantly the abrupt growth is insensitive to wave number $k$, as evidenced from the same abrupt growth for $k=0.1 k_c$ mode and $k=10 k_c$ mode in the case of $\theta_0=179^{o}$ in Fig. (\ref{fig: compare_passive_full_axion_90_179}).  This provides a crucial clue for the growth mechanism.

The restoring forces of long-wave modes and short-wave modes are very different, with the former being negative and the latter being positive.  Hence the same growth for all $k$ modes indicates that the cause of the growth should be from the source, the right-hand side of Eq. (\ref{equ: perturbed field equation}).   Unlike the free-particle model, well before the onset of nonlinear mass oscillation the source is almost zero, where $\theta'\to 0$ and $m^2a^2\sin(\theta)\to 0$ as $\theta\to\pi$.  The weak constant source yields a small coefficient in the $a^6$ growth (phase (a)) before the abrupt growth, more so for $\theta_0$ closer to the potential top, as opposed to a much larger coefficient due to a much larger source in the $\theta_0=\pi/2$ case.  Just at the onset the source suddenly rises to its full strength when the field $\theta$ rolls down the hill on its first pass.  Such a drive is so abrupt that the perturbed field gets amplified regardless of the nature of its restoring force, since the restoring force has no time to respond.   One may analogize this mechanism as the "direct current (DC)" drive, as opposed to the "alternative current (AC)" drive of the parametric instability to be discussed in the next section.  After this short period of time, the source strength either stays in full strength or declines depending on whether the mode has entered horizon.  For super-horizon modes, the source stays in full strength and the modes enter phase (c) of a slow $a^2$ growth, and for sub-horizon modes, they enter a new phase of parametric instability or matter-wave oscillation, phase (d).

The exact location of $\theta_0$ from the top would, however, affect the onset time of mass oscillation.   For a given $k$, the more delay of the mass oscillation, the longer the duration of the $a^6$ growth, and the perturbation can grow to a greater amplitude.   On the other hand, the closer $\theta_0$ is to the field potential top, the smaller is the source, and the smaller the coefficient of the $a^6$ growth as mentioned above.  These two opposite trends almost cancel, and by the end of the abrupt growth, $\delta\theta$ is brought to nearly the same amplitude as the free-particle model.   An alternative way to understand this is that once the source becomes at its full strength, it drives the perturbed field to a level comparable to the photon perturbation before the perturbed field becomes decoupling from the source shortly after the onset of mass oscillation.   Such a driving mechanism applies to all adiabatic perturbations.

\section{Parametric Instability}
\label{sec: Parametric Instability}

Parametric instability refers to the presence of an oscillating restoring force of almost twice the natural frequency for an oscillator, described by Mathieu's equation \cite{AS1964}:
\begin{equation}
\label{equ: Mathiu's equation}
\ddot{Q} + \omega_\theta^2[1+\eta+\epsilon\cos(2\omega_\theta t)]Q=0,
\end{equation}
where $Q$ is the oscillator solution and the overdot denotes $d/dt$.   The parameters $\epsilon$ and $\eta$ are the driver strength and the detuning squared frequency.  The $(\eta, \epsilon)$ phase diagram at small $\epsilon$ and $\eta$ marks the marginally stability curve as $|\eta| = (1/2)|\epsilon|$.  In the limit $\eta =0$, the oscillator is unconditionally unstable even for an tiny but finite $\epsilon$.

To make a comparison with Mathieu's equation, we change the variable of Eq. (\ref{equ: perturbed field equation}) from the scaling factor $a$ to the ordinary time $t$.  A straightforward algebra shows that Eq. (\ref{equ: perturbed field equation}), Taylor-expanded up to the first-order nonlinearity in $\theta$, can be cast into the equation:
\begin{equation}
\label{equ: expansion of the perturbed field equation}
\left.\begin{aligned}
& \ddot{q} + \Big \{ {3\over 4}\mathscr{H}^{2} + {{k^2}\over{a^2}} + m^2 \Big [1-{{\langle\theta^2\rangle}\over 4}\Big ( 1 + \cos \Big (2\int_{t_m}^t\omega_\theta dt\Big ) \Big ) \Big ] \Big \}q \\
& = {Sr\over{a^{{1\over2}}}},
\end{aligned}
\right.
\end{equation}
where $q\equiv \delta\theta/\sqrt{\epsilon_\theta}$ with $\epsilon_\theta$ being the background energy density, $\mathscr{H}=H/a$, $t_m$ is the time for the onset of mass oscillation, $\omega_\theta$ the frequency of $\theta$ containing a nonlinear frequency shift, and $Sr$ is the right-hand side of Eq. (\ref{equ: perturbed field equation}).  The short-time average $\langle\theta^2\rangle$ decays as $a^{-3}$, which we model as $\langle\theta^2\rangle=(\theta_0^2/2) (t/t_m)^{-3/2}$.

The frequency $\omega_\theta$ of the nonlinear oscillation of $\theta$ can be derived from Eq. (\ref{equ: background field equation}), where the restoring force $\sin(\theta)=\theta(1-\theta^2/6)+O(\theta^5) \sim \theta_0(t/t_m)^{-3/4}[1-(t/t_m)^{-3/2}(\theta_0^2/8)]\cos(\int \omega_\theta dt))-[\theta_0^3(t/t_m)^{-9/4}/24]\cos(3\int \omega_\theta dt)$, assuming $\theta=\theta_0(t/t_m)^{-3/4}\cos(\int\omega_\theta dt)$.  Ignoring the triple frequency term and retaining the coefficient of $\cos(\int\omega_\theta dt)$, we have the driving frequency
\begin{equation}
\label{equ: omega_theta formula}
\omega_\theta^2 = m^2 \Big [ 1-{{\theta_0^2}\over8}\Big ( {t\over{t_m}} \Big )^{-{3\over2}} \Big ].
\end{equation}

On the other hand, the perturbed field $q$ has a natural frequency $\omega_\delta$ different from the driving frequecy $\omega_\theta$ and related by
\begin{equation}
\label{equ: omega_delta formula}
\omega_\delta^2 = \omega_\theta^2\Big [ 1 - {{\theta_0^2}\over8}\Big ( {t\over{t_m}} \Big )^{-{3\over2}} \Big ],
\end{equation}
to the leading order.

We shall address the sub-horizon regime where $k /a \gg \mathscr{H}$.  Hence we can ignore both the weak source term $Sr$ as the driver $\phi$ declines as $a^{-2}$ shown in Paper (\Rmnum{1}) and the $\mathscr{H}^2$ term in Eq. (\ref{equ: expansion of the perturbed field equation}), thus arriving at a simplified equation that describes the homogeneous solution of $q$,
\begin{equation}
\label{equ: simplified expansion of the perturbed field equation}
\left.\begin{aligned}
& \ddot{q} + \Big \{ {{k^2}\over{a^2}} + \omega_\theta^2\Big [ 1-{{\theta_0^2}\over 8}\Big ({t\over{t_m}} \Big )^{-{3\over2}}\Big ( 1 + 2\beta\cos\Big (2 \int_{t_m}^t \omega_\theta dt \Big ) \Big ) \Big ] \Big \} q \\
& = 0,
\end{aligned}
\right.
\end{equation}
with $\beta=1$.  An addtional parameter $\beta$ is introduced so as to make a close contact with the Mathieu's equation which has two parameters $\epsilon$ and $\eta$.

Now, Eq. (\ref{equ: simplified expansion of the perturbed field equation}) is the Mathieu's equation with time-dependent coefficients, where the detuning squared frequency $\eta = (k^2/a^2)-\delta\omega^2$ and the driver strength $\epsilon= 2\delta\omega^2$ with $\delta\omega^2\equiv \omega_\theta^2(\theta_0^2/8)(t/t_m)^{-3/2}$. Note that when $(k/a)^2\to 0$, we have $|\eta| = (1/2)|\epsilon|$ and it satisfies the marginally stable condition of Mathieu's equation.   Worth noting is that the squared detuning frequency $\eta$ has zero-crossing for a range of $k$ and $a$, and these $k$-modes can temporarily be parametrically unstable.   This provides a crude explanation as to why in some range of $k$ the matter-wave oscillation appears to be amplified and has a relatively high amplitude, i.e., feature (3), but more details will follow.

Other than the aforementioned growth due to the parametric drive, the frequency of the solution of Eq. (\ref{equ: simplified expansion of the perturbed field equation}) actually deviates from its natural frequency and is locked to near half of the driving frequency for some period; therefore the solution becomes nearly phase locked to the driver during this period.  As shown in Paper (\Rmnum{1}), we may let $q= \Re[\hat q \exp(-i\int \omega_\theta dt)]= \Re[\hat q]\cos(\int\omega_\theta dt) +\Im[\hat q]\sin(\int\omega_\theta dt)]$ while the background field $\theta=\sqrt{\epsilon_\theta}\cos(\int\omega_\theta dt)$, where $\hat q$ is a slowly varying complex amplitude.   When $q$ and $\theta$ are phase locked, the amplitudes of $\Re[\hat q]$ and $\Im[\hat q]$ will remain fixed and do not oscillate until the nonlinearity dies out, after which the perturbation assumes free-particle matter-wave oscillation.  This picture provides a rough baseline as to why the first half cycle of matter-wave oscillation in $\Delta_\theta$ has a long duration.  Again, more details are to come.

We shall analyze an even more simplified version of Eq. (\ref{equ: simplified expansion of the perturbed field equation}) below, which bears more resemblance to Mathieu's equation, in order to bring out the aforementioned frequency locking and the amplitude excess in a quantitative manner.  We assume the background field oscillates at a fixed frequency, $\omega_\theta = m$, ignoring the nonlinear contribution to the driving frequency which is a high-order effect for our purpose.  Equation (\ref{equ: simplified expansion of the perturbed field equation}) thus becomes
\begin{equation}
\label{equ: more simplified expansion of the perturbed field equation}
\left.\begin{aligned}
& \ddot{q} + \Big \{ {{k^2}\over{a^2}} + \omega_\theta^2\Big [ 1-{\theta_0^2\over8}\Big ( {t\over{t_m}} \Big )^{-{3\over2}}\Big ( 1 + 2\beta\cos(2\omega_\theta (t-t_m)) \Big ) \Big ] \Big \}q \\
& = 0,
\end{aligned}
\right.
\end{equation}

Using this $\hat q$ representation for sub-horizon modes after mass oscillation, one can show that the normalized energy density $\Delta_\theta\approx 2\Re[\hat q]$.  Aside from the coefficient, the interaction ($\beta$) term in Eq. (\ref{equ: more simplified expansion of the perturbed field equation}) yields $(\Re[\hat q]/2)[\cos(\omega_\theta (t-t_m))+\cos(3\omega_\theta (t-t_m))] -(\Im[\hat q]/2)[\sin(\omega_\theta (t-t_m)) - \sin(3\omega_\theta (t-t_m))]$.  Again ignoring the triple frequency contribution, the interaction term is then proportional to $(1/2)[\Re[\hat q]\cos(\omega_\theta (t-t_m)) - \Im[\hat q]\sin(\omega_\theta (t-t_m))]=(1/2)\Re[\hat q^*\exp(-i\omega_\theta(t-t_m))]$.  Substituting this result into Eq. (\ref{equ: more simplified expansion of the perturbed field equation}), we have a reduced perturbation equation satisfying
\begin{equation}
\label{equ: amplitude equation}
i\dot{\hat q} = {1\over{2\omega_\theta}}\Big [\Big(\Big ( {k\over a} \Big )^2-\alpha \Big )\hat q  - \beta\alpha\hat q^*\Big],
\end{equation}
where $\alpha =\omega_\theta^2(\theta_0^2/8)(a_m/a)^3$, and $a_m$ is the scaling factor at the onset of nonlinear mass oscillation. Separating the real and imaginary parts of $\hat q$, one can straightforwardly show that the dispersion relation for this equation is $\omega=(1/2\omega_\theta)[ [ (k/a)^2 -\alpha ]^2 - (\beta\alpha)^2 ]^{1/2}$ with $\omega$ being the matter-wave frequency.  This dispersion relation yields the characteristics of parametric instability.  For the $\beta=1$ axion case, the dispersion relation becomes
\begin{equation}
\label{equ: dispersion ralation0}
\omega = {k \over 2\omega_\theta a}\Big[\Big({k \over a}\Big)^2-2\alpha\Big]^{1/2},
\end{equation}
and the mode is unstable when $k^2/2a^2\alpha < 1$, and stable with $\omega \to k^2/2\omega_\theta a^2$ when $k^2/2a^2\alpha \gg 1$.  This dispersion relation is valid even when $\omega_\theta\neq m$ where Eq. (\ref{equ: omega_theta formula}) holds.  For simplicity we shall continue to ignore the nonlinear correction to the driving frequency and assume $\omega_\theta = m$.

We first note the factor $k^2/(2a^2\alpha) \propto k^2(a/a_m)(a_m)^{-2}$, the greater $a_m$ of nonlinear mass oscillation is, or the closer $\theta_0$ is to $\pi$, the smaller the magnitude of this factor at a given $a/a_m$, thus mimicking a smaller $k$ for the free-particle model that has a longer matter-wave oscillation period and accounts for feature (2).  On the other hand, for a given $\theta_0$ near $\pi$ and in the limit $k\to 0$, the parametric instability is weak, as by the time when the mode enters horizon where Eq. (\ref{equ: dispersion ralation0}) becomes valid, the nonlinearity is already small.  So the only range of $k$ exhibiting a strong parametric growth is when $k$ is on the same order of $k_c$.  This explains why the amplitude excess occurs for $k$ on the same order of $k_c$, feature (3).

To put the above into quantitative perspectives, one readily sees from the dispersion relation, Eq. (\ref{equ: dispersion ralation0}), that the frequency's being either imaginary and small compared with the free-particle frequency is to contribute to higher perturbation amplitudes and a longer duration in the first half cycle of matter-wave oscillation.  The unstable phase takes place during a time interval $\ln(a_0/a_{m0})=2\ln(\theta_0/2)+3\ln(a_m/a_{m0})+2\ln(k_c/k)]$ or $\ln(a_0/a_k) = 2\ln(\theta_0/2)+3\ln(a_m/a_{m0})+\ln(k_c/k)$ depending on whether the mode has entered horizon or not, respectively, at the onset of nonlinear mass oscillation, where $a_0$ is the scaling factor at the end of growth $\omega=0$, and $a_k$ that at the horizon entry $k=2H$.  Here $a_{m0}$ corresponds to $a$ at the onset of free-particle mass oscillation, i.e., $2H(a_{m0})=ma_{m0}$, and the critical wave number for the free particle model, $k_c = (2m a H)^{1/2}=m a_{m0}$; by the same token, we have $a_k/a_{m0}=k_c/k$ using the definition of horizon entry that $a_k k=2aH$.  (The quantity $aH=a_{m0} H(a_{m0})$ since it is redshift-independent in the radiation era, and we thus have $k_c \propto m^{1/2}\propto a_{m0}^{-1}$.)

As a supplementary remark, the above estimate for the duration of unstable phase has taken into account that prior to this parametric growth, low-$k$ super-horizon mode must go through the $a^2$ growth of phase (c) even after the onset of nonlinear mass oscillation, where the driving source $Sr$ is still strong and Eq. (\ref{equ: simplified expansion of the perturbed field equation}) is not valid; for such modes, only after horizon entry, $a=a_k$, does Eq. (\ref{equ: simplified expansion of the perturbed field equation}) become valid and hence the solution of this equation starts at $a_k$.

For these sub-horizon $k$ modes subject to parametric instabilities, the amplitudes increase by a growth factor proportional to $\exp[A(a_\kappa)-A(a)]$, and the exponent of the growth factor at the end of parametric growth can be shown to be
\begin{equation}
\label{equ: maximum growth}
\left.\begin{aligned}
A(a_\kappa) = {{2k^2}\over{k_c^2}}\Big \{ & \Big [ {{a_{m0}}\over {a_\kappa}}\Big ( {{a_m}\over{a_{m0}}} \Big )^3{{k_c^2}\over{k^2}}{{\pi^2}\over4}-1 \Big ]^{1\over2} - \\
                                             &  \tan^{-1}\Big [ \Big ( {{a_{m0}}\over {a_\kappa}}\Big ( {{a_m}\over{a_{m0}}} \Big )^3 {{k_c^2}\over{k^2}}{{\pi^2}\over4}-1 \Big )^{1\over2} \Big ] \Big \},
\end{aligned}
\right.
\end{equation}
using WKB approximation, where $a_\kappa=a_k$ and $a_\kappa=a_m$ correspond to $a_k > a_m$ and $a_k < a_m$, respectively, $A(a_0)$ is defined to be $0$, and $\theta_0$ takes the value $\pi$. (See Appendix for derivation.)   This growth factor is responsible for the power excess.  When $(a_m/a_\kappa)[(a_m/a_{m0})(k_c/ k)(\pi/2)]^2 \gg 1$, the growth factor $\exp[A(a_\kappa)]$ becomes $\exp[\pi (k/k_c)(a_m/a_{m0})^{3/2}(a_{m0}/ a_\kappa)^{1/2}]$.  Therefore for long waves we have small amplification, $A(a_k)\propto (k/k_c)^{3/2}$ as $a_{m0}/a_k = k/k_c$.   On the opposite limit for short waves, $k > k_c (a_m/a_{m0}) (\pi/2)$, the parametric growth would never occur.  This explains why we do not see the power excess for long waves and short waves in Fig. (\ref{fig: compare_passive_full_axion_90_179}).

This unstable phase is followed by a matter-wave oscillation phase but with a lower frequency than normal.  The solution in this phase has a form $\sin(B(\eta) + \pi/4)$, where the detail is also given in Appendix using WKB approximation, and we have
\begin{equation}
\label{equ: phase_B formula}
\left.\begin{aligned}
B(\eta) = 2{{k^2}\over {k_c^2}}\Big \{& -[1-r\exp(-\eta)]^{1/2} \\
                                      & +{1\over 2}\ln \Big [ {{1+[1-r\exp(-\eta)]^{1\over2}}\over {1-[1-r\exp(-\eta)]^{1\over2}}} \Big ] \Big \}.
\end{aligned}
\right.
\end{equation}
Here $\eta\equiv\ln(a/a_m)$ and $r=(k_c/k)^2(\pi^2/4)(a_m/a_{m0})^2$.  This oscillation has an initial $B(\eta)=0$ at $\eta=\eta_0$ where $\omega=0$, but otherwise $B(\eta) \ge 0$ for $\eta \ge \eta_0$.

The peak of the power excess (feature 3) should be located in this oscillation phase since the solution $\sin(B(\eta)+\pi/4)$ is still on the rise at $\eta_0$.  If one is to assume that solutions have resumed free-particle matter-wave oscillations when they reach the peaks, i.e., $r\exp[-\eta]\ll 1$, then one may find the timing $\eta_{peak}(\equiv\ln(a_{peak}/a_{m0}))$ of the solution peaks as a function of $k$, $m$ and nonlinearity given by $\sin[B(\eta_{peak}) +\pi/4]=1$, or $B(\eta_{peak}) +\pi/4=\pi/2$.   We thus have
\begin{equation}
\label{equ: eta_peak}
\left.\begin{aligned}
\eta_{peak} = & {\pi\over 4}\Big({{k_c} \over k}\Big)^2+2(1-\ln(2))+ \\
              & \Big [ 2\ln\Big ( {{k_c}\over{k}} \Big )+3\ln \Big ( {{a_m}\over{a_{m0}}} \Big )+2\ln\Big ( {\pi\over2} \Big ) \Big ],
\end{aligned}
\right.
\end{equation}
where terms in the squared bracket are contributed from the growing phase.

Subtracting $\eta_{peak0}$ of the free-particle model from this $\eta_{peak}$ , we can determine the total delay in the first quarter cycle of nonlinear mass oscllation.  As shown in Appendix (A) of Paper (\Rmnum{1}), the free-particle model has a oscillating solution $\propto\sin[(k/k_c)^2[\eta-\ln((\sqrt{3}/2)(a_\kappa/a_{m0})+c_0(k)]]$ in phase (d).  Here, the phase $c_0(k)=\cos^{-1}[1/((1+(k/k_c)(\gamma-0.5))^2)^{1/2}]$ with $\gamma$ being the Euler number $\sim 0.577$, and thus $c_0(k)$ is nearly $0$ for a wide range of $k/k_c$.  Now, using $(k/k_c)^2(\eta_{peak0}-\ln((\sqrt{3}/2)(a_\kappa/a_{m0}))= \pi/2$ to fix $\eta_{peak0}$, we obtain the total delay $\Delta\eta(=\eta_{peak}-\eta_{peak0})$ as
\begin{equation}
\label{equ: Delta_eta}
\left.\begin{aligned}
\Delta\eta_{peak} \approx & -{\pi\over 4}\Big({{k_c}\over k}\Big)^2 + (1+\kappa)\ln \Big ( {{k_c}\over {k}} \Big )+3\ln \Big ( {{a_m}\over {a_{m0}}} \Big ) \\
                   & +2+2\ln \Big ({\pi\over 4} \Big )+\ln\Big({2\over\sqrt{3}}\Big),
\end{aligned}
\right.
\end{equation}
where $a_k/a_{m0} =k_c/k$ has been used, and $\kappa=0$ for long waves where $a_\kappa = a_k$ and $\kappa = 1$ for short waves where $a_\kappa=a_{m0}$.

This is an interesting prediction, in that the strong negative $(k_c/k)^2$ dependence of $\Delta\eta_{peak}$ can make the delay be negative.  The cause of the reverse effect is that the growing phase of parametric instability brings the amplitude to $1/\sqrt{2}$ of the peak in a time weakly dependent on $k_c/k$.   This period can be short compared to the free matter-wave oscillation to bring the amplitude to a similar level for long waves, which take a time $\delta\eta\sim (4/\pi)(k_c/k)^2$.  The maximum $\Delta\eta_{peak}$ can be found by taking a derivative of it with respect to $k_c/k$ and the maximum delay is found to be near $k/k_c\sim 1$.  This explains why the delay in the first half cycle of nonlinear mass oscillation is prominent only around $k\sim k_c$ in Fig. (\ref{fig: compare_passive_full_axion_90_179}).

Finally, since our results above depend on $a_m$, it is useful to pin down the relation between $\delta\theta_0(\equiv\pi-\theta_0)$ and $a_m$.   One can Taylor expand the field potential gradient near $\theta_0=\pi$ in Eq. (\ref{equ: background field equation}) where $\sin(\theta) \approx \theta-\pi=\delta\theta$.   Since the results, Eqs. (\ref{equ: maximum growth}), (\ref{equ: eta_peak}) and (\ref{equ: Delta_eta}), are derived using Mathieu's equation, Eq. (\ref{equ: more simplified expansion of the perturbed field equation}), to be consistent with these results, one should define $a_m$ in accordance with the power-law-amplitude assumption.  We extrapolate the asymptotic power-law solution, $\langle\theta^2\rangle \propto t^{-3/2}$, backward in time till it intercepts the actual background solution at $\pi^2$ to define the onset time $t_m$.   In so doing, we find the following analytical formula provides the best fit:
\begin{equation}
\label{equ: a_m versus deltatheta_0}
\Big ( {{a_m} \over {a_{m0}}} \Big )^2\approx 3.5 - {2\over3}\ln\Big ( {{\delta\theta_0}\over{\delta\theta_{01}}} \Big ),
\end{equation}
where $\delta\theta_{01}=1^{o}$.  This expression works fairly fine; using it to compute $\eta_{peak}$ of Eq. (\ref{equ: eta_peak}) gives $<10\%$ errors against the measured $\eta_{peak}$.

\section{Numerical Solution and General Nonlinear Model}
\label{sec: Numerical Solution and General Nonlinear Model}

\begin{figure}
\includegraphics[scale=0.35, angle=270]{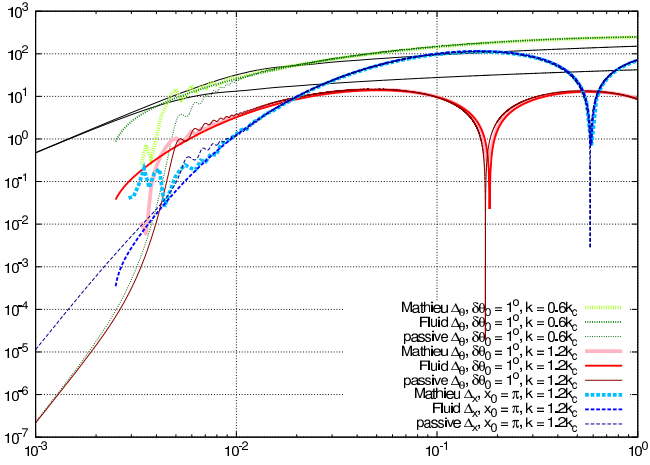}
\caption{Comparison of $\Delta_\theta$'s constructed from passive evolution, Mathieu's equation, Eq. (\ref{equ: simplified expansion of the perturbed field equation}) and the fluid equation, Eq. (\ref{equ: delta n equation}).  The solutions given by Mathieu's equation and fluid equation agree very well with those of passive evolution long after the onset of nonlinear mass oscillation.  Plotted here are also solutions of passive evolution for another nonlinear scalar field model with a potential $\propto 1-\sech(x)$ and of the corresponding Mathieu's equation and fluid equation for comparison, and excellent agreement is also found. Black lines are passive CDM perturbations for $k=0.6$ and $1.2 k_c$.  Particle mass $m=10^{-22}$ eV is assumed.}
\label{fig: compare_passive_parametric}
\end{figure}

In Fig. (\ref{fig: compare_passive_parametric}), we plot $\Delta_\theta$'s constructed from the numerical solutions of Eq. (\ref{equ: perturbed field equation}) for passive evolution.  The fiducial particle mass $m=10^{-22}$ eV is chosen.   A comparison to $\Delta_\theta$'s constructed from Eq. (\ref{equ: simplified expansion of the perturbed field equation}) is also shown here, where we take $\beta=1$ and the metric fluctuation $\phi=0$ (c.f., Eq. (\ref{equ: definition of Delta psi})); the initial solution slope is set to $\dot q/q= [\omega_\theta^2(3\pi^2/8-1)-k^2/a_m^2]^{1/2}$ (c.f., Eq. (\ref{equ: simplified expansion of the perturbed field equation})).  Clearly seen in Fig. (\ref{fig: compare_passive_parametric}) is good agreement between the two solutions, except in the early time where our leading-order Taylor expansion of the nonlinearity fails.  This plot demonstrates that peculiar features (2) and (3) appearing in the solution of Eq. (\ref{equ: perturbed field equation}) indeed arise from the parametric drive.  These two features show strongly for $k\sim k_c$ modes than for $k\ll k_c$ modes as explained in the last section.  In Fig. (\ref{fig: compare_passive_parametric}), we also plot a third solution of a fluid equation derived from the Gross-Pitaevskii equation, which is the subject of the next section, and in Appendix, we present the fluid equation, Eq. (\ref{equ: delta n equation}).  This fluid equation filters out the high-freqeuncy mass oscillation and is therefore an equation for the slowly varying amplitude.  One can see that this third solution agrees with the solution of Mathieu's equation, Eq. (\ref{equ: simplified expansion of the perturbed field equation}), extremely well.

\begin{figure}
\includegraphics[scale=0.35, angle=270]{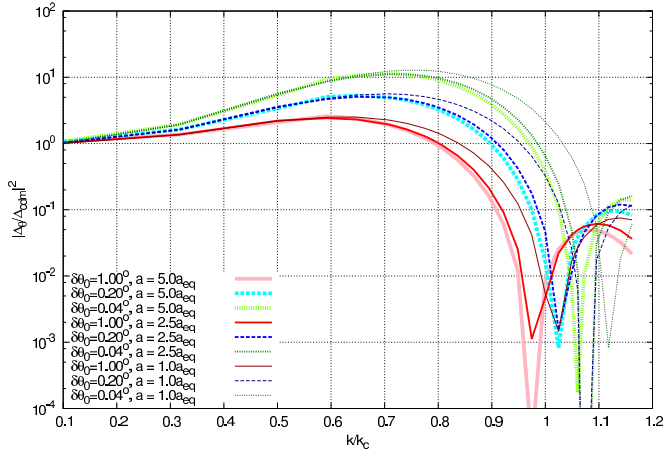}
\caption{Transfer functions of the extreme axion model of $m=10^{-22}$ eV with different initial angles at $a_{eq}$, $2.5 a_{eq}$ and $5a_{eq}$. Note that the quantity $k_{peak}$ is almost frozen ever since $a=a_{eq}$. The three initial angles correspond to the axion field strengths, $4\pi G f^2 = 1.71$, $1.13$, $0.821\times 10^{-5}$ from large to small $\theta_0$.}
\label{fig: transfer_function}
\end{figure}

The matter spectra not long after the radiation-matter equality is particularly interesting since by then the primary spectral feature can hardly evolve and is frozen throughout the later evolution.   We plot in Fig. (\ref{fig: transfer_function}) the transfer function, $|\Delta_\theta/\Delta_{cdm}|^2$, of several initial field angles $\Delta\theta_0$'s at $a=a_{eq}, 2.5a_{eq}$ and $5a_{eq}$ using the full treatment, where $a_{eq}$ is the scaling factor at the radiation-matter equality.   One can clearly see the broad spectral bumps in all extreme initial angles.   For $k$ at and smaller than the spectral peak the transfer function barely evolves after $a_{eq}$, but are opposite for larger $k$.  This is due to the Jeans wave number around $a_{eq}$ is close to $k_c$ [Paper (\Rmnum{1})];  above the Jeans wavelength, perturbations grow self-similarly as those of CDM but below, perturbations become neutrally stable oscillating matter waves \cite{Hu2000, WooChiueh2009}.

In evolving toward $a=5a_{eq}$, the photon-electron decoupling occurs around $a=3a_{eq}$ and the photon perturbation contributes very little to the metric perturbation since then.  In Appendix B of Paper (\Rmnum{1}), we showed that the drag between baryons and photons already damps out the photon perturbation prior to photon-electron decoupling for $k > k_c$, and hence the metric perturbation indeed has no contribution from photons regardless of whether or not we have considered the electron-baryon recombination physics.  But for $k \ll k_c$, the drag is ineffective and hence metric perturbations are affected by photons at $a > 3a_{eq}$ if the silk damping is not properly accounted for, and can produce  some errors in the matter power spectrum. Our full treatment does not take in to account the silk damping.  However, this error for matter perturbations is practically small since matters are cold and photons are hot, and gravity responds to cold matter. Moreover, these small errors in both CDM and axion perturbations are the same in our full treatment since at long waves the two perturbations are almost identical.  Hence the transfer function is insensitive to such errors present in their respective spectra.   It is based on this rationale the transfer functions at $a=5a_{eq}$ are presented in Fig. (\ref{fig: transfer_function}).

In Fig. (\ref{fig: transfer_function}), we have made sure that the dark matter energy density $m^2f^2\theta^2$, together with baryon energy density, equals the radiation energy density at $a_{eq}$.  As $\theta_0$ approaches $\pi$ that delays the onset of nonlinear mass oscillation, the value of $f$ must decrease to satisfy the above condition.  Normally the field strength is characterized by a dimensionless parameter $f^2/m_p^2(\equiv 4\pi G f^2)$, where $m_p$ is the Planck mass.  The free-particle case corresponds to $f^2/m_p^2\to \infty$.  The appropriate parameter regime for the extreme axion model has values $f^2/m^2\sim O(10^{-5})$, and so $f$ is on the order of GUT scale.

To further demonstrate the general validity of Eq. (\ref{equ: simplified expansion of the perturbed field equation}) approximating the original perturbed field equation,  we consider the potential $(m^2a^2)[1-\sech(x)]$, where $x$ is the field.  We replace mass terms in the field equations, Eqs. (\ref{equ: background field equation}) and (\ref{equ: perturbed field equation}), by $(m^2a^2)\tanh(x)\sech(x)$ and $(m^2a^2)\sech(x)[-1+2\tanh^2(x)]\delta x$, respectively, which yield $\beta=1$, $\alpha=(5x_0^2/8)(t/t_m)^{-3/2}$ and $\omega_x^2=m^2-\alpha$.  Here we also choose the initial field value $x_0=\pi$.  Plotted also in Fig. (\ref{fig: compare_passive_parametric}) is the comparison of $\Delta_x$'s constructed from the passive evolution and Eq. (\ref{equ: simplified expansion of the perturbed field equation}).  Again, excellent agreement is found when $t \gg t_m$, reinforcing our claim for the parametric drive of the original perturbed field equation.

To end this section, we notice that $\beta=1$ is quite generic to all symmetric field potentials, and this can be shown as follows.  Let the Taylor expansion of the field potential be $V(x) = (m^2a^2/2)(x^2-(b/2)x^4+O(x^6))$, and the potential gradient $V^{'}=(ma)^2(1-bx^2)x\sim m^2(1-(3b/4)x_0^2)x\equiv\omega_x^2 x$, where $x_0$ is the oscillation amplitude, and $\omega_x$ is the nonlinear driving frequency adopting the technique used for the axion case.  The coefficient of the restoring force in the perturbation equation is $V^{''} = (ma)^2(1-3bx^2)$, which can be reduced to $\omega_x^2[1 -(3b/4)x_0^2(1+ 2\cos(2\omega_x t))]$.  In Sec. (\ref{sec: Parametric Instability}), we have parametrized the last factor as $(1+2\beta\cos(2\omega_x t))$, and hence $\beta =1$ for all nonlinear models with symmetric potentials with a finite mass.

\section{Axion connection to Gross-Pitaevskii Equation}
\label{sec: Axion connection to Gross-Pitaevskii Equation}

The Gross-Pitaevskii equation reads
\begin{equation}
\label{equ: Gross-Pitaevskii equation}
i{{\partial}\over{\partial t}}\Phi = \Big [ -{{\nabla^2}\over{2m}} + g|\Phi|^2 \Big ]\Phi.
\end{equation}
This non-relativistic equation can be deduced from the Bogoliubov's formalism in the zero-temperature limit of a dilute interacting Bose-Einstein condensate \cite{Bogoliubov1947, Pitaevskii2013}.  The interaction energy density is proportional to $g|\Phi|^4$, where $g$ is a coupling constant related to the microscopic scattering length $a_0$ as $g\equiv(4\pi/m) a_0$.  The dilute gas approximation demands that $|a_0|^3 |\Phi|^2 \ll 1$, and the interactions are repulsive (attractive) when the coupling constant $g$, or the scattering length $a_0$, is positive (negative).

The dynamics of the linear perturbation $\delta\Phi =\Phi-\Phi_0$, with the background field $\Phi_0$ chosen to be real, can be best seen using a fluid approach.   The field $\Phi$ is expressed in the polar coordinate as $\Phi = \xi\exp[i\chi]$, the fluid density $n\equiv \xi^2$ and the fluid velocity ${\bf v}\equiv\nabla \chi/m$, the quantum specific pressure $-(\nabla^2 \xi)/(2m\xi)$ and the fluid specific pressure $gn$.   The total specific pressure is $-\nabla^2 \xi/(2m\xi) + gn$, and when $g$ is negative the total specific pressure can become negative in the long-wave limit and it is straightforward to show that the dispersion relation is $\omega^2 = (k^2/2m)[k^2/(2m) + 2mgn]$. (See Appendix.)   If we identify $\alpha = -2mgn$ with a negative $g$ and $k/a$ replaces $k$, this dispersion relation is identical to Eq. (\ref{equ: dispersion ralation0}) for the axion model.

To compare with Mathieu's equation Eq. (\ref{equ: Mathiu's equation}) in detail, we examine the excitations of the Gross-Pitaevskii equation, Eq. (\ref{equ: Gross-Pitaevskii equation}).  Note that the background filed ($k=0$) has a frequency $\omega_0 = g|\Phi_0|^2$, and $\delta(|\Phi|^2\Phi) = 2|\Phi_0|(\Re[\Phi_0]\Re[\delta\Phi]+\Im[\Phi_0]\Im[\delta\Phi])+|\Phi_0|^2 \delta\Phi$.  We can remove the above complication by defining a frequency-shift wave function $p = \Phi\exp[i g|\Phi_0|^2 t]$ for all modes including the background field, then $\delta(|\Phi|^2 \Phi)= \exp[-ig|\Phi_0|^2 t]|\Phi_0|^2(2\delta p + \delta p^*)$.    Therefore the linearized equation becomes
\begin{equation}
\label{equ: linearlized Gross-Pitaevskii equation}
 i\dot\delta p+g|\Phi_0|^2 \delta p = \Big [ {{k^2}\over{2m}} + 2g|\Phi_0|^2 \Big ] \delta p + g|\Phi_0|^2 \delta p^*.
\end{equation}
This equation is identical to Eq. (\ref{equ: amplitude equation}) for the axion model ($\beta=1$) after replacement of coefficients: $\alpha=-2mg|\Phi_0|^2$ and $a=1$. Thus, aside from the source due to metric perturbations, the linear perturbations of Eq. (\ref{equ: expansion of the perturbed field equation}) and the excitations of Gross-Pitaevskii's equation are the same.  Not only that, the background field of Eq. (\ref{equ: expansion of the perturbed field equation}) and the ground state of the Gross-Pitaevskii equation are also the same.  The ground state is uniform with a frequency $g|\Phi_0|^2$, and the background field is also uniform having a frequency $m-(\alpha/2m)$.  Subtracting off the leading order mass oscillation frequency to get to the non-relativistic regime, we find the ground state and the background field oscillating at the same frequency.  Therefore the axion model can therefore be fit into Bogoliubov's framework of dilute interacting boson gas.   (We have noted that there have been previous works on the connection between the perturbed scalar-field equation, Eq. (\ref{equ: perturbed field equation}), and the Gross-Pitaevskii equation, but unfortunately in a rather ad hoc manner where the nonlinear shift of the driving frequency was not taken into account thus yielding a higher nonlinear coupling strength $g$ by a factor $12$ than it actually is \cite{MM2012}.)

The interaction potential energy for axion, to the leading order, is about $(mf)^2\theta^4 = (m/f)^2\Psi^4$, and so $|g|=f^{-2}$, since the mass (energy) density $2m^2\langle\Psi^2\rangle = m|\Phi|^2$.  It implies that the scattering length $|a_0|= m/(4\pi f^2)$, and the dilute gas condition, $|a_0|^3|\Phi|^2 \sim (4\pi)^{-3} (m/f)^4\ll 1$, valid to an excellent degree.  However, for all practical purposes, this naive estimation gives too small a $|a_0|^3 |\Phi|^2$ by many orders of magnitude for $m$ as small as $10^{-22}$eV, due to the fact that the scattering length $|a_0|$ is many orders of magnitude smaller than the Planck length.    This problem arises from the fact that the gravity has been ignored in Bogoliubov's formulation.  If the scattering length $|a|$ is limited to the smallest possible length, the Planck length $l_{p}$, the dilute gas condition becomes $l_p^3f^2 m$ and is still much less than unity even when $f$ takes the largest possible value, the Planck scale $l_p^{-1}$, therefore justifying the dilute gas approximation.

Having the microscopic physics in place, we can now ask how reasonable it is for the initial angle so close to the top of the field potential, as quantum tunneling may have made its way to render the system unstable.   Note that the macroscopic axion field $\Psi = \sqrt{N}\psi_q$, where $\psi_q$ is the quantum field of individual particle and $N$ is the number of particles overlapped in some macroscopic coherent length.    The background field has an infinite coherent length in the zero-temperature limit.   Let us take a conserved position where the perturbed field has a coherent length comparable to one wavelength $2\pi/k$; the number $N$ is still a huge number due to the extremely small particle mass.  As an example, the fiducial mode $k=k_c\sim 10$ Mpc$^{-1}$ encloses a particle number $N=n_m \lambda_c^3\sim 10^{98}$, where $n_m$ is the number density at the onset of free-particle mass oscillation and $\lambda_c$ is the Compton wave length of the fiducial particle mass $m_{22}=10^{-22}$ eV. Therefore the particle number amounts to $N \sim 10^{98}(m/m_{22})^{-5/2}(k/k_c)^{-3}$. The quantum tunneling is suppressed by the factor $\prod_i\exp[-S_i/\hbar]$, where $S_i$ is the Eulerian action of an individual particle $i$.   The suppression factor $\exp[-S_i/\hbar]$ is usually taken to be a Gaussian around the classical field $\psi_c$.   Though individual $S_i/\hbar$ can be small enough to permit quantum tunneling, especially near a classical bifurcation point, the coherent N particles share the same phase space coordinates, thus $S_i=S$, and the suppression factor becomes $\exp[-NS/\hbar]$.  It greatly narrows the variance of the Gaussian around the classical field.  Hence, tunneling through quantum fluctuations is impossible even when the initial field angle is very close to the potential top.

In Bogoliubov theory, there is another quantity, the healing length $l_h$ defined to be $(2m|g||\Phi|^2)^{1/2}l_h \equiv 1$, which characterizes the relative strength of destabilizing nonlinear to stabilizing linear terms of Eq. (\ref{equ: simplified expansion of the perturbed field equation}).  In the absence of self-gravity, structures beyond $l_h$ grow due to the weak instability of negative $g$ until nonlinear structures, such as vortex filaments \cite{CC2006}, form on the scale of $l_h$.  But with self-gravity, smaller nonlinear structures can form through much stronger gravitational instabilities.  The healing length $l_h$ is about the Compton wavelength at the onset of mass oscillation, and since $l_h \propto a^{3/2}$ in physical coordinate, it becomes about $150$ Mpc in the present universe for particle mass $m=10^{-22}$ eV.  It is interesting to note that this healing length coincides with the observed correlation length of the baryon acoustic oscillation at present \cite{Eisenstein2005}.   Will the healing length have cosmological footprints on very large-scale structures?  As the baryon acoustic oscillation scale depends linearly on $a$ and the healing length on $a^{3/2}$, future observations at high redshift will tell.

\section{Conclusion}
\label{sec: Conclusion}

In this work, we analyze the three unexpected features of the extreme axion and explain their underlying mechanisms.  Among them, the parametric drive and amplification mechanism accounts for two non-trivial features.  To illustrate of the mechanisms, we show the original perturbed field equation can be made equivalent to Mathieu's equation, which is able to faithfully recover the two features.  We also disclose that in the non-relativistic limit and to the leading-order nonlinearity, the equation of motion for the axion model is identical to the Gross-Pitaevskii equation, a macroscopic manifestation of a zero-temperature interacting Bose-Einstein condensate.  Based on this connection, we explain why the quantum tunneling of this system is impossible.

The two nontrivial features, i.e., extension of spectral cutoff to higher wave number and spectral excess of the extreme axion model can have important impacts in structure formation of the high-$z$, matter-dominant universe, due to the fact that most first-generation galaxies formed out of perturbations near the spectral cutoff.  The spectral cutoff is determined solely by particle mass in the free-particle model.  Its extension to higher $k$ for the extreme axion model mimics the effect of higher particles mass for free particle.  Therefore, the high-$k$ power spectrum may not be a good indicator accurately reflecting the true particle mass in the extreme axion model.  Recent simulations addressing the high-redshift Lyman-$\alpha$ absorption features indicate that substantially higher particle mass than $10^{-22}$eV is required or implied \cite{APYNB2017,IVHBB2017}.  On the other hand, approximately $10^{-22}$eV particle mass is needed to account for the flat cores of dwarf spheroidal galaxies \cite{Schive2014, CSC2017}.  The tension in particle mass may be lessened with the extreme axion model.

However we have found a limit to the high-$k$ spectral extensions, no matter how extreme a condition the initial angle assumes.  The spectral extensions are all confined to wave numbers less than a factor $2$ higher from that of the free-particle model, i.e., approximately $< k_c$.  That is, the spectral excess peaks around $0.6- 0.8 k_c$ and immediately following the spectral peak is a sharp cutoff.  This spectral shape renders the first collapsed halo of mass $[(4\pi/3)(\pi/k)^3]\rho_0\sim 10^{10} M_\odot$, where $k$ is near the peak of the spectral excess and $\rho_0$ is the background mass density.  As a reference, the first galaxies in the free-particle axion model of $m=10^{-22}$ eV have masses several $\times 10^{10} M_\odot$ \cite{Schive2016}.

The spectral excess is perhaps our most surprising finding, since conventional dark matter candidates proposed so far are unable to produce power excess over the CDM model across the perturbation spectrum.  When $\delta\theta_0 < 1$ degree, the spectral excess can be so distinct that may completely revise the standard scenario of first galaxy formation.  First of all, the spectral excess leads to earlier formation of first generation galaxies and push the reionization epoch \cite{Planck2016} earlier than the free-particle model \cite{Schive2016}.   Second, taking the more extreme case $\theta_0=0.2$ degree as an example, c.f., Fig. (\ref{fig: transfer_function}), the broad spectral peak yields first collapsed halo of mass $10^9-10^{10} M_\odot$, and frequent mergers of these over-abundant first halos than the conventional are to quickly build up more massive halos.   Furthermore, busy mergers are prone to sustain intense star bursts and even rapid super-massive black hole growths.   Finding quasar at $z=7$ \cite{Mortlock2011} and recent discovery of galaxies more massive than the Milky Way inferred to already form at $z=7$ \cite{Glazebrook2017} have posed challenges for the CDM model.   Given the aforementioned possible outcomes, the extreme axion model may stand a better chance to meet such a challenges.  Whether successful or not, only future simulations can tell.

To place our results in a concrete ground, we provide a formula for the wave number of spectral peak $k_{peak}$ and a procedure for the peak height to be calculated, as functions of the initial angle and the particle mass in Appendix.  Aside from that, our identification of the perturbed field equation to the Gross-Pitaevskii equation in the non-relativistic limit is of practical relevance.  It permits calculations of pertubation dynamics using a fluid approach outlined in the beginning of Sec. (\ref{sec: Axion connection to Gross-Pitaevskii Equation}) and carried out in Appendix.

Finally, we must stress that the extreme axion model is not a special model capable of producing the three peculiar features studied in this work.  A wide range of scalar field models have the same characteristics, as demonstrated in Sec. (\ref{sec: Numerical Solution and General Nonlinear Model}) by an example. In any of these scalar field models, the particle mass $m$ has to be extremely light, not far from our fiducial mass $10^{-22}$ eV, to produce astronomical observable effects.  As to the nonlinear effect of general scalar field models, if the coupling constant $g$ in the Gross-Pitaevskii equation, proportional to the microscopic scattering length, increases with time as the scaling factor $a$, then the parametric instability explored in this work will have a long-lasting, but weak, effect for perturbations beyond a particular length scale.

\bibliography{Reference}

\renewcommand{\theequation}{A\arabic{equation}}
\setcounter{equation}{0}
\appendix
\renewcommand{\theequation}{A\arabic{equation}}
\section{Particle Mass Dependence}
\label{app: Particle Mass Dependence}

In Sec. (\ref{sec: Parametric Instability}) we consider how solutions change with a changing $a_m$ for a fixed particle mass $m$, and concluded that when $a_m$ becomes large, the effect is equivalent to make $k$ appear smaller.  When the particle mass $m$ is allowed to change, Eq. (\ref{equ: amplitude equation}) is invariant to a transformation, which we call the mass-tempo-size trasformation.  We normalize time $t$ to $t_m$, and note that $\alpha\propto m^2(a_m/a)^3$.  It is straightforward to show that the solution is invariant, up to a shift in $\ln(a)$ space, to the changing $a_m$, $k$ and $m$ so long as $k a_m$ and $m a_m^2$ are kept fixed.  This transformation for a changing $a_m$ is in fact a transformation only of the particle mass $m$.  We note that $\ln(a_m) = \ln(a_{m0}) + \ln K(\delta\theta_0)$ for a $K(\delta\theta_0)$ given in Eq. (\ref{equ: a_m versus deltatheta_0}); the quantity $\ln(a_{m0})$ depends only on the particle mass $m$ and another quantity $K(\delta\theta_0)$ on how nonlinear the background field is, and hence the particle mass and the nonlinearity appear to be able to vary independently.  However, the transformation requires a fixed $ma_m^2$, which becomes $m a_{m0}^2 K^2(\delta\theta_0) = (aH) K^2(\delta\theta_0)$ where $aH$ is independent of the particle mass $m$, the scaling factor $a$ and the nonlinearity.  As a result, a fixed $m a_m^2$ implies fixed nonlinearity $K(\delta\theta_0)$ in the tranformation, and therefore this transformation involves only a varying particle mass $m$.

Realizing it, the $m$ dependence of the spectral peak $k_{peak}$ can be straightforwardly obtained.   This can be carried out since an analytical solution of the perturbed field can be found to a good approximation.  Adopting the fluid approach to the Gross-Pitaevskii equation, Eq. (\ref{equ: Gross-Pitaevskii equation}),  in the comoving frame with proper replacement of $\nabla \to i k/a$ and others explained in last section, Sec. (\ref{sec: Axion connection to Gross-Pitaevskii Equation}), we find that the sub-horzon normalized fluid density $\delta n/n$ satisfies
\begin{equation}
\label{equ: delta n equation}
\left.\begin{aligned}
& {{d^2}\over{d\eta^2}}\Big ( {{\delta n}\over n} \Big ) + \Big ( {k\over{k_c}} \Big )^4\Big [ 1-\Big ( {{k_c}\over k} \Big )^2{{\theta_0^2}\over4}\Big ({{a_m}\over{a_{m0}}} \Big)^2\exp(-\eta)\Big ] \Big ( {{\delta n}\over n} \Big ) \\
& =0,
\end{aligned}
\right.
\end{equation}
where $\eta\equiv \ln(a/a_{m})$ and $\delta n/n = 2\Re[\hat q]\approx \Delta_\theta$. 
  
As a check of the accuracy of this fluid equation to Eq.(\ref{equ: simplified expansion of the perturbed field equation}), we solve Eq.(\ref{equ: delta n equation}) numerically.  The initial $\eta_i$ starts from $a_\kappa/a_{m0}$ with the initial slope $S(\eta_i)\equiv (k/k_c)^2 [1-(k_c/k)^2(\theta_0^2/4)(a_m/a_{m0})^2 \exp[-\eta_i]]^{1/2}$. Solutions are plotted in Fig. (\ref{fig: compare_passive_parametric}), and one can see that the solutions excellently agree with those of Mathieu's equation.

One may adopt the WKB approximation to analyze the solution, for which the phase, $\int d\eta S(\eta)$, has an analytical expression.  When the integrand is imaginary, it represents a growing solution $\exp(A(\eta_i)-A(\eta))$, as $A(\eta)$ is a decreasing function of $\eta$.  When the integrand is real, it has an oscillating solution, $\sin(B(\eta)+d_0)$, where $B(\eta)\geq 0$ and is an increasing function of $eta$, and $d_0$ is the phase. In between, the integrand crosses zero, the WKB approximation fails and we have an Airy function that connects solutions on the two sides, i.e., $\exp[A_(\eta_i)-A(\eta)] \to 2\sin(B(\eta)+\pi/4)$.  Here, the analytical expressions of $A$ and $B$ are
\begin{equation}
\label{equ: formula of A and B}
\left.\begin{aligned}
& A(\eta) = \\
& 2{{k^2}\over {k_c^2}} \Big \{ [r\exp(-\eta)-1]^{1\over2}-\tan^{-1}[((r\exp(-\eta)-1)^{1\over 2}] \Big \}, \\
& B(\eta) = \\
& 2{{k^2}\over {k_c^2}}\Big \{-[1-r\exp(-\eta)]^{1\over 2}+{1\over 2}\ln \Big [ {{1+[1-r\exp(-\eta)]^{1\over2}}\over {1-[1-r\exp(-\eta)]^{1\over2}}} \Big ] \Big \},
\end{aligned}
\right.
\end{equation}
where $r=(k_c/k)^2(\pi^2/4)(a_m/a_{m0})^2$, and both $A(\eta)$ and $B(\eta)$ equal zero when the integrand crosses zero.   The expression of $A(\eta)$ has been used in Sec.(\ref{sec: Parametric Instability}) to find the growth factor.

When $\eta\gg 1$, we recover that free-particle oscillation, i.e., $B \to (k^2/k_c^2)(\ln(a/a_m)+ d)$ with a phase $d$. Note that the peak of the solution is located at the oscillating side of the solution.  The relation between $k_{peak}$ and $m$ is just the solution of a transcendental equation $B(\eta_{eq})+\pi/4 =\pi/2$, where the sine oscillation phase is equal to $\pi/2$, where $\eta_{eq}$ is the value of $\eta$ at radiation-matter equality.  We make a further approximation to simplfy the matter.  The nonlinear contribution to $B$ is negligible at the peak of the solution, i.e., $r\exp^(-\eta) \ll 1$, so that $B(\eta)\approx (k/k_c)^2(\ln(a/a_{m0})+[\ln(a_{m0}/a_m)+2(\ln(2)-1)-\ln(r)])$.  This approximation is better for a large $m$ than for a small $m$.    We thus obtain a simpler transcentdental equation for $k_{peak}$:
\begin{equation}
\label{equ: k_peak versus a_eq}
\left.\begin{aligned}
& {\pi\over 4}\Big ( {{k_c}\over {k_{peak}}} \Big )^{2} + \ln \Big [{\pi\over 4}\Big ( {{k_c}\over{k_{peak}}} \Big )^2 \Big ] \\
& =\ln \Big ( {{a_{eq}}\over{a_{m0}}} \Big )-3\ln \Big ( {{a_m}\over{a_{m0}}} \Big )-2-\ln\Big ( {\pi\over4} \Big ),
\end{aligned}
\right.
\end{equation}
which can be solved numerically rather easily.  It can be easily seen that $k_{peak}/k_c <1$ for $m=10^{-22}$ eV, and more so for a larger particle mass because $\ln(a_{eq}/a_{m0})$ gets larger.   Aside from the mass dependence of $k_c\propto m^{1/2}$, $k_{peak}$ has another weak mass dependence on $\ln(a_{m0}/a_{eq})$, with $a_{m0}\propto m^{-1/2}$.  Another term, $\ln(a_m/a_{m0})$, is a measure of $\delta\theta_0$, which is given in Eq. (\ref{equ: a_m versus deltatheta_0}) and has no mass dependence.

Though $k_{peak}$ is derived here from the passive evolution, $k_{peak}$ of the full treatment deviates only slightly from this formula; thus, to a good approximation Eq. (\ref{equ: k_peak versus a_eq}) provides an analytical expression for $k_{peak}$ of the full treatment, and we find this expression is accurate within $10\%$ of the peak of $k^3|\Delta_{\theta}(k)|^2$ of the full treatment.  Moreover, $k_{peak}$ is largely frozen after $a=a_{eq}$ shown in Fig. (\ref{fig: transfer_function}), as $k_{peak}$ is smaller than the Jeans wave number $k_J$ in the matter-dominated regime, and therefore this spectral peak persists in the linear matter power spectrum throughout the later epoch.

The particle mass dependence of the quantity $k_{peak}/k_c$ is mild, shown in the right-hand side of Eq. (\ref{equ: k_peak versus a_eq}).  When the particle mass $m\to \infty$, we have $\ln(a_{eq}/a_{m0})\to 0$, the peak $k_{peak}/k_c\to 0$, and the growth factor of Eq. (\ref{equ: maximum growth}) approaches zero.   The particle mass scaling of the growth factor can be calculated by determining $k_{peak}$ and substituting into the growth factor Eq. (\ref{equ: maximum growth}).  Comparing the growth factor for $m=10^{-22}$ eV to find the ratio, one is then able to determine the spectral peak height for any $m$ by referring to the peak height of Fig. (\ref{fig: transfer_function}) for $m=10^{-22}$ eV, which can be approximated to be $\ln|\Delta_\theta/\Delta_{cdm}|^2|_{peak} = 17(k_{peak}/k_c) - 10.2$ as a fit.

\label{lastpage}

\end{document}